\documentclass[12pt]{article}

 \newread\testifexists
 \def\GetIfExists #1 {\immediate\openin\testifexists=#1
     \ifeof\testifexists\immediate\closein\testifexists\else
     \immediate\closein\testifexists\input #1\fi}

 \usepackage{gthstyle}\usepackage{amsfonts}
 \usepackage{amssymb}
 \immediate\openin\testifexists=e
     \ifeof\testifexists\immediate\closein\testifexists\else
     \immediate\closein\testifexists\input e\fipsf

 \def\Bbb#1{\setbox0=\hbox{$\tt #1$}  \copy0\kern-\wd0\kern .1em\copy0}

 \def\bbf#1{\setbox0=\hbox{$#1$} \kern-.025em\copy0\kern-\wd0
         \kern.05em\copy0\kern-\wd0 \kern-.025em\raise.0433em\box0}

 \immediate\openin\testifexists=a
     \ifeof\testifexists\immediate\closein\testifexists\else
     \immediate\closein\testifexists\input a\fimssym.def  

 \def\a{\alpha}      \def\b{\beta}         
 \def\d{\delta}         
        \def\L{\Lambda}     \def\m{\mu}
 \def\f{\phi}                \def\n{\nu}
             \def\r{\varrho}       
 \def\t{\tau}          
               
 \def\w{\omega}

 \def\pa{\partial} \def\ra{\rightarrow}
  
 \def\dd{{\rm d}}  \def\bra{\langle}   \def\ket{\rangle}

 \def\fract#1#2{{\textstyle{#1\over#2}}}
 \def\ffract#1#2{\raise .2 em\hbox{$\scriptstyle#1$}\kern-.3em/
                 \kern-.2em\lower .15 em \hbox{$\scriptstyle#2$}}

 \def\part#1#2{{\partial#1\over\partial#2}}

                     \newcommand{\fn}{\footnote}
 \newcommand{\nn}{\nonumber\\[2pt]}             
 \newcommand{\be}{\begin{eqnarray}}             \newcommand{\ee}{\end{eqnarray}}
 \newcommand{\bi}[1]{\begin{itemize}\item[#1]}         
       \newcommand{\ei}{\end{itemize}}
 \newcommand{\eqn}[1]{(\ref{#1})}

 \newcommand{\crlb}[1]{\label{#1}\\[2pt]}
 \newcommand{\eela}[1]{\quad\hbox{\scriptsize{#1}}\label{#1}\end{eqnarray}}
 \newcommand{\eelb}[1]{\label{#1}\end{eqnarray}}
 
 \newcommand{\newsecb}[2]{\section{#1}\label{#2}\setcounter{equation}{0}}
 
 \newcommand{\nolabels} {\def\eel{\eelb} \def\crl{\crlb} \def\newsecl{\newsecb}}

\newcommand\publishversion{\nolabels\setlength{\textheight}{9in}\setlength{\oddsidemargin}{0in}
    \setlength{\textwidth}{6.3in}\setlength{\topmargin}{-0.1in}}

\publishversion
\begin{document} \begin{titlepage}

\title{\normalsize \hfill ITP-UU-09/38  \\ \hfill SPIN-09/35
\\ \vskip 20mm \Large\bf
Quantum Gravity without Space-time Singularities or Horizons\thanks{Erice School of Subnuclear
Physics 2009}}

\author{Gerard 't~Hooft}
\date{\normalsize Institute for Theoretical Physics \\
Utrecht University \\ and
\medskip \\ Spinoza Institute \\ Postbox 80.195 \\ 3508 TD
Utrecht, the Netherlands \smallskip \\ e-mail: \tt g.thooft@uu.nl \\ internet: \tt
http://www.phys.uu.nl/\~{}thooft/}

\maketitle

\begin{quotation} \noindent {\large\bf Abstract } \medskip \\
In an attempt to re-establish space-time as an essential frame for formulating quantum gravity -- rather than an ``emergent"
one --, we find that exact invariance under scale transformations is an essential new ingredient for such a theory. Use is
made of the principle of ``black hole complementarity", the notion that observers entering a black hole describe its
dynamics in a way that appears to be fundamentally different from the description by an outside observer. These differences
can be boiled down to conformal transformations. If we add these to our set of symmetry transformations, black holes,
space-time singularities, and horizons disappear, while causality and locality may survive as important principles for
quantum gravity.
\end{quotation}

\vfill \flushleft{\small{Version September 18, 2009}}

\end{titlepage}

\eject

\newsecl{Introduction: scales in general relativity}{intro}

Due to quantum effects, a black hole emits particles of all sorts. Thus, if left entirely by itself, a black hole gradually
looses mass, eventually ending its life with a gigantic explosion. This is the Hawking effect\cite{Hawking}. For an
intuitive understanding of our world, the Hawking effect seems to be quite welcome. It appears to imply that black holes are
just like ordinary forms of matter: they absorb and emit things, they have a finite temperature, and they have a finite
lifetime. One would have to admit that there are still important aspects of their internal dynamics that are not yet quite
understood, but this could perhaps be considered to be of later concern. Important conclusions could already be drawn: the
Hawking effect implies that black holes come in a denumerable set of distinct quantum states. This also adds to a useful and
attractive picture of what the dynamical properties of space, time and matter may be like at the Planck scale: black holes
seem to be a natural extension of the spectrum of elementary physical objects, which starts from photons, neutrinos,
electrons and all other elementary particles.

In such a picture, however, what happens to the horizon and the space-time singularities? An answer sometimes suggested by
string theorists, as well as others, is that all of space-time is just
``emergent"\cite{emergent}\cite{graphiti}\cite{girelli}; the theory should first be formulated without space-time
altogether. Or, perhaps, time alone is an emergent concept\cite{Isham}. It was argued that, at least, locality would have to
be abandoned\cite{giddings}\cite{giddingsetal}. In this paper, however, we dismiss all such options. In particular, we
insist that any satisfactory theory should have built in a strong form of causality, as well as locality, in order to
explain why cause precedes effect, and why events separated at some distance from one another appear to evolve
independently. For this, space-time appears to be indispensable.

Something has to give, and in this paper we claim to have found a good candidate for that: the definition of \emph{scales}
in space-time. It should be done in a way that differs from conventional wisdom.

In what follows it is useful to write the metric tensor \(g_{\m\n}(x)\) as:
 \be \matrix{g_{\m\n}(x)&=\ \w^2(x)\,\hat{g}_{\m\n}(x)\ ,&\hbox{where}\qquad\nn
 \det(\hat{g}_{\m\n})&=\ -1\ ,\qquad\w\ =&(-\det(g_{\m\n}))^{1/8}\ .}\eel{scaledmetric}
Since the light cone is defined by solutions of the equation
 \be\dd s^2=g_{\m\n}\dd x^\m\dd x^\n\ ,\eel{lightcone}
it is \(\hat{g}_{\m\n}\) that determines the positions of light cones, whereas \(\w(x)\) defines the scales for rulers and
clocks in the theory.

As for general coordinate transformations, the split \eqn{scaledmetric} simply implies that \(\hat g_{\mu\nu}(x)\) and
\(\w(x)\) transform anomalously. When \(x^\m\ra u^\m(x)\), then
 \be \matrix{\hat g_{\mu\nu}(x)&\ra& \tilde{\hat g}_{\mu\nu}(u)&=&{\pa x^\a\over \pa u^\mu}{\pa x^\b\over \pa u^\nu}\,\hat
 g_{\a\b}(x)\,\det\left({\pa x\over\pa u}\right)^{-1/2}\ ;\cr
\w(x)&\ra& \tilde\w(u)&=&\w(x)\,\det\left({\pa x\over\pa u}\right)^{1/4\vphantom{]^|}}\ .\qquad\quad} \eel{ghattransform}

\newsecl{Black hole complementarity}{compl}
An observer entering a black hole would experience quantized fields \(\f(x)\) that are defined in a space-time expanding
across the horizon. Indeed, this observer would not even know where exactly the horizon is, and her description should
certainly not depend on that. Assuming the horizon to be a plane that we put at the origin of our coordinate frame, we can
regard space-time as to be split into two pieces:
 \be\hbox{region } I\ ,& z>0,\ \f(x)=\f_I(\r,\t)\ ,&\hbox{with}\quad z=\r\cosh\t\ ,
 \crl{regionone}
 \hbox{region } II\ , & z<0,\ \f(x)=\f_{II}(\r,\t)\ ,&\hbox{with}\quad z=-\r\cosh\t\ .\eel{regiontwo}
So here, \(\f(x)\) is the field used by the ingoing observer, while the outside observer has only \(\f_I\) to his disposal.
One derives that the vacuum state \(|\hbox{vac}\ket\) experienced by the ingoing observer, splits up as
 \be |\hbox{vac}\ket=\sum_n|n\ket_I|n\ket_{II}e^{-\pi E_n}\ , \eel{rindlersplit}
where \(|n\ket_I\) is the complete set of states of the outside observer in region \(I\), while
\(|n\ket_{II}\) is a complete set of states that all escape the attention of this observer. \(E_n\)
are the energies in these states, and \(\t\) is the normalized time coordinate for the outside
observer. The probability that a state \(n\) is detected is therefore
 \be W(n)=C\,e^{-2\pi E_n}\ , \eel{probn} so that we have a temperature \(T\) which, in the units
 used, obeys \(kT=1/2\pi\).

\emph{Black hole complementarity}\cite{GtHcompl1}\cite{GtHcompl2}\cite{GtHcompl3}\cite{susskind} refers to the fact that
observers who stay outside the black hole can see the Hawking particles, including the effect they have on the space-time
metric, while observers entering the black hole can neither see these particles, nor the effect they have on the metric. On
the other hand, these observers do have to add all particles that entered, and will enter, the black hole to obtain a
meaningful description of what is going on there. These differences are caused by the fundamental difference in the way the
two sets of observers experience the notion of time.

We now bring a slight twist in this notion of complementarity. Rather than distinguishing observers
who go into a black hole from the ones staying out, we note that the ingoing observer sees
everything that went into the black hole, whereas the second observer sees all objects going out --
the Hawking radiation. Thus reformulated, complementarity really refers to the mapping from the
in-states to the out-states.

It is instructive then, to consider the special case that the black hole was formed by the collapse of a single shell of
massless non interacting particles moving in radially, because this case has a simple exact analytical solution: outside the
shell we have the Schwarzschild metric, inside the shell space-time is flat, and the two regions are glued together in such
a way that the metric is continuous. In particular, the transverse components of the metric must match, which means that the
\(r\) coordinate coincides inside and outside the shell. Furthermore, we take the case that the outgoing Hawking particles
also form a single shell of matter. Since, actually, the Hawking particles have a thermal spectrum, it is highly unlikely
that these particles emerge as a single shell, but it is not impossible. The shell is suppressed by large Boltzmann factors,
but this just means that it occupies a very tiny portion of phase space; that is not relevant to our procedure. Later, one
may replace the in- and outgoing matter configurations by more realistic ones. Thus, for simplicity, the in-configuration is
described by Fig.~\ref{bhshell.fig}$a$, and the out-configuration by \ref{bhshell.fig}$b$.

\begin{figure}[h] \setcounter{figure}{0}
\begin{quotation}
 \epsfxsize=120 mm\epsfbox{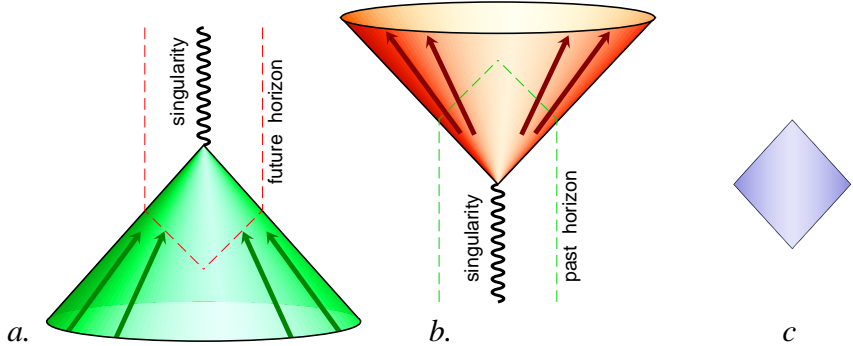}
  \caption{$a)$ The in-configuration of a black hole; $b)$ the out-configuration. Dotted lines are
  the horizons in both pictures. In the case both the in-going matter and the out-going matter forms a single shell,
  there is a region at the center that both pictures seem to have in common ($c$).}
  \label{bhshell.fig}\end{quotation}
\end{figure}

According to the complementarity principle, the ingoing observer thinks that only the in-going matter in
Fig.~\ref{bhshell.fig}$a$ is responsible for the matter contribution to the space-time curvature, while the Hawking
radiation is essentially invisible. In Fig.~\ref{bhshell.fig}$b$, space-time is shown as it is recorded by the late outside
observer: \emph{only} the Hawking particles are responsible for the space-time curvature, while all in-going matter that
formed the black hole in turn is invisible to him\fn{This picture differs somewhat from the older versions of
``complementarity", where the outside observer does see the matter going in; we presently take the viewpoint that if an
observer registers the effect of Hawking radiation on the metric, all ingoing matter becomes invisible, which is the exact
time-reverse of what the ingoing observer sees.}. Naturally, one asks how these differences can be implemented in one
all-embracing theory. Are there transformations from one picture to the other? How complex may these transformations be? How
can these be reconciled with causality?

\newsecl{Causality and locality}{causal}

We now insist that the complementarity transformation that relates Fig.~\ref{bhshell.fig}$a$ with Fig.~\ref{bhshell.fig}$b$
should be a local transformation of dynamical physical degrees of freedom, and furthermore, that this transformation
preserve the causal order of events, so that an evolution law obeying obvious causality can be phrased in either one of the
two systems. This necessarily means that the light cones should not be affected by this transformation. Therefore, in
Eqs.~\eqn{scaledmetric} and \eqn{lightcone}, \(\hat g_{\mu\nu}(x)\) should remain the same, but \(\w(x)\) may be different.
To be precise, \(\w(x)\) is observer dependent. The value of \(\w(x)\) depends on whether the point \(x\) is seen through a
curtain of Hawking particles or directly. Therefore, according to the complementarity principle, the physical events taking
place at some space-time point \(x\) should not depend explicitly on \(\w\). We end up with a picture where \(\w(x)\) is
entirely unobservable locally, much like a local gauge parameter \(\L(x)\) in a gauge theory.

The fields \(\hat g_{\mu\nu}(x)\) must obey equations of their own. Formulating these equations
will be difficult, and we postpone this to future work. By first studying the physical consequences
of this idea, we must cultivate some feeling about how to proceed.

At this point, we can already identify some of these consequences. Suppose we have a space-time where we are free to choose
\(\w\). Suppose that \emph{cosmic censorship}\cite{cosmiccensor}\cite{Wald} holds. This means that, in terms of the
conventional metric \(g_{\mu\nu}\), any naked singularity is hidden behind a horizon. Then, whenever we encounter such a
singularity, we can adjust \(\w\) in such a way that a clock shows infinite time when approaching this singularity.
Therefore, it now occurs at \(t=\infty\). In practical examples, the singularity will also have been smeared out so that it
effectively disappears, and then also the horizon disappears, as we will demonstrate.

\newsecl{Conformal transformations}{conformal}
In the case described in Section~\ref{compl}, where both in- and out-going matter forms a single shell, there is an
internal flat region that both descriptions have in common. Since it is flat, they have the same, flat value \(\hat
g_{\mu\nu}(x)=\eta_{\mu\nu}\) for the metric. So here, the only choice we have for the transformation is the value for
the scales \(\w(x)\). As is well-known, the only transformations that leave \(\hat g_{\mu\nu}=\eta_{\mu\nu}\) intact,
are the conformal transformations. If in both Figures~\ref{bhshell.fig}$a$ and \ref{bhshell.fig}$b$, we choose the tip
where the singularity starts as the origin of space-time, then this mapping is
\def\inn{\mathrm{in}}\def\outt{\mathrm{out}}
 \be x^\mu_\inn =(-1)^{\displaystyle \d^{\mu o}}{x^\mu_\outt\over |(x_\outt )^2|}\ , \eel{conftrf}
where the square is of course the Lorentz invariant square of the coordinate \(x_\outt\). We chose
the sign here such that the timelike causal order is preserved. All vectors in the relevant region
are timelike with respect to the origin chosen.
\begin{figure}[h]
\begin{quotation}
 \epsfxsize=120 mm\epsfbox{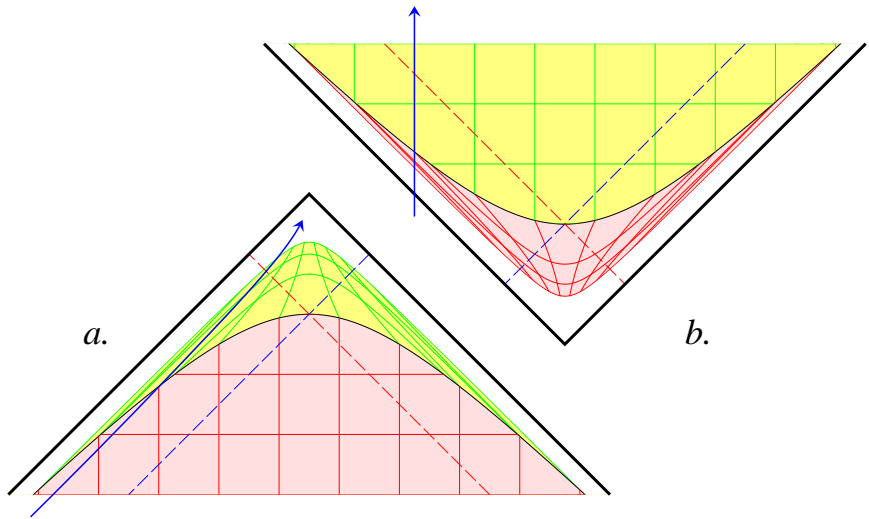}
  \caption{The conformal transformation for the inside of the black hole of Fig.~\ref{bhshell.fig}.
Bold lines show matter going in ($a$), and out ($b$). The dotted lines show how lightlike geodesics transform.
The arrowed line is the same timelike trajectory in the two frames.}
  \label{conformal.fig}\end{quotation}
\end{figure}

The transformation \eqn{conftrf} is illustrated in Figure~\ref{conformal.fig}, where we see that indeed lightlike geodesics
stay lightlike. We see cosmic censorship at work: the singularity in both frames are moved to \(t\ra\pm\infty\) in the other
frames. As for the region outside the collapsing and the expanding shells, we might decide to choose no change in \(\w\). At
the lightlike shells themselves, the mapping is singular, but this may be attributed to the fact that we are dealing with
macroscopic black holes in this example, so that, in both coordinate frames, matter tends to take divergent values for their
stress-energy-momentum tensors. In microscopic black holes, the transformation may be less singular at these regions.

\newsecl{Scale invariance and the stress-energy-momentum tensor}{scaleinv}
The above considerations imply that the scale factor \(\w(x)\) is ambiguous, yet it is needed to
describe clocks and rulers in the macroscopic world, and it is also needed if we wish to compute
the Riemann and the Ricci curvature, because they depend on the entire metric \(g_{\mu\nu}\), not
just \(\hat g_{\mu\nu}\). Clearly, the world that is familiar to us is not scale invariant. Without
\(\w\) we cannot define distances and we cannot define matter, but we can define the geometry of
the light cones, and, if \(\hat g_{\mu\nu}\) is non-trivial, it should describe some of the
physical phenomena that are taking place. It should be possible to write down equations for it
without directly referring to \(\w\).

To what extent will \(\w(x)\) be determined by \(\hat g_{\mu\nu}(x)\), if we impose some physical conditions? For instance,
we can impose the condition that the space-time singularities are all moved to \(t\ra\infty\), so that the interiors of
black holes at the moment of collapse, are re-interpreted as the exteriors of decaying black holes, as in
Figs.~\ref{bhshell.fig} and \ref{conformal.fig}. After that is done, conformal transformations are no longer possible,
because they would move infinities from the boundary back to finite points in space-time, which we do not allow. But we do
wish to impose some extra ``gauge" condition on \(\w(x)\), so as to fix its value, allowing us to define what clocks,
rulers, and matter are.

A natural thing to impose is that all particles appear to be as light as possible. If we would have
only non-interacting, massless particles, then \(T_\m^\m=0\). Therefore, a natural gauge to choose
is the vanishing of the induced Ricci scalar:
 \be R(\,\w^2\hat g_{\mu\nu})=0\ . \eel{Riccinul}
When \(\hat g_{\mu\nu}(x)\) is given, we can compute (in 4 space-time dimensions):
 \be &&\matrix{R_{\mu\nu} &=&\hat R_{\mu\nu} +\w^{-2}(4\,\pa_\mu\w\,\pa_\nu\w-\hat g_{\mu\nu}\hat
 g^{\a\b}\,\pa_\a\w\,\pa_\b\w)\cr &&+\ \vphantom{\int^{[k]}}\w^{-1}(-2D_\mu\,\pa_\nu\w-\hat g_{\mu\nu}\hat g^{\a\b}D_\a\,\pa_\b\w)\
 ;}\crl{riccitensor}
&&\w^2\,R \ =\ \hat R -6\,\w^{-1}\hat g^{\mu\nu}D_\mu\pa_\nu\w
 \ . \eel{ricciscalar}
Here, we write \(R_{\mu\nu}(\w^2\hat g)=R_{\mu\nu}\), and \(R_{\mu\nu}(\hat g_{\mu\nu})=\hat
R_{\mu\nu}\), and similarly the Ricci scalars. The covariant derivative \(D_\mu\) treats the field
\(\w\) as a scalar with respect to the metric \(\hat g_{\mu\nu}\).

 The condition that Eq.~\eqn{ricciscalar} vanishes gives us the equation
 \be  \hat g^{\mu\nu}D_\mu\pa_\nu\w(x)=\fract16\hat R \,\w(x)\ . \eel{riccicondition}
This happens to be a linear equation for \(\w(x)\), which is the reason for having included the square in its
definition, in Eq.~\eqn{scaledmetric}, since \(\w\) is not restricted to be infinitesimal.

Now let \(\w(x)\) obey Eq.~\eqn{riccicondition} in a region where \(\hat g_{\mu\nu}=\eta_{\mu\nu}\), so that \(\hat R=0\),
and consider the Fourier transform \(\w(k)\) of \(\w(x)\), which, at \(k\ne 0\), we now take to be infinitesimal. The wave
vector \(k\) obeys \(k^2=0\). Let this wave vector be in the \(+\) direction. Then, according to Eq.~\eqn{riccitensor}, the
only non-vanishing component of the Ricci tensor, and hence also the only non-vanishing component of the
stress-energy-momentum tensor, is the \(++\) component. Thus, we have shells of massless particles traveling in the
\(x^-\)-direction. Adding all possible Fourier components, we see that, in the infinitesimal case, our transition to
non-trivial \(\w(x)\) values leads us from the vacuum configuration to the case that massless particles are flying around.
As long as \(\w(k)\) at \(k\ne 0\) stays infinitesimal, these massless particles are non-interacting.

In the case of an evaporating black hole, these massless particles are the Hawking particles. Our complementarity
transformation, the transformation that modifies the values of \(\w(x)\) while keeping Eq.~\eqn{riccicondition} valid,
switches on and off the effects these Hawking particles have on the metric\fn{It is often claimed that Hawking particles
have little or no effect on the stress-energy-momentum operator on the horizon\cite{Howard}\cite{HowardCandelas}. This,
however, is true only for the expectation values in the states used by the ingoing observer (the Hartle-Hawking states). The
outgoing observer uses Boulware states, for which the \(T_{\mu\nu}\) diverges\cite{Jensen}\cite{Visser}.}.

\begin{figure}[h]
\begin{quotation}
 \epsfxsize=100 mm\epsfbox{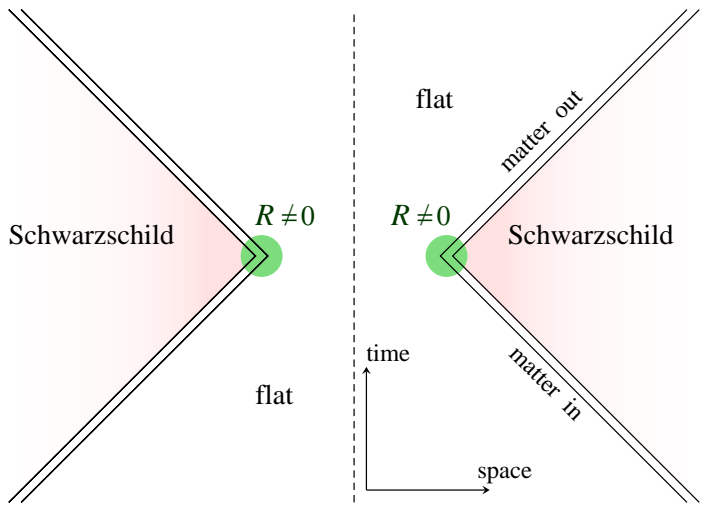}
  \caption{Combining the different conformal frames into one metric for the imploding and subsequently evaporating
  black hole, gives us a region of large Ricci scalar values \(R\) near the horizon (here, the two circles). Note that
  the metric outside the collapsing and evaporating shells is Schwarzschild, while inside the shells it is flat.
  Demanding \(\w\) to be continuous across the shells implies that the \(r\) coordinate should be matched there (see
  Ref.~\cite{DraytH}).}
  \label{inoutmetric.fig}\end{quotation}
\end{figure}

Often, it will seem to be more convenient to impose that we have flat space-time, \emph{i.e.} \(\w^2\hat
g_{\mu\nu}=\eta_{\mu\nu}\) at infinity, rather than imposing \(R=0\) everywhere. In the case of an implosion followed by an
evaporating black hole (for simplicity emitting a single shell of matter as in Section~\ref{compl}), this would allow us to
glue the different conformal regions together, to obtain Figure~\ref{inoutmetric.fig}. This then gives us a region of large
Ricci scalar curvature near the horizon. Ricci scalar curvature is associated with a large burst of pressure in the matter
stress-energy-momentum tensor. This means that an observer who uses this frame sees a large explosion near the horizon that
sends the ingoing material back out, that is, the black hole explodes classically. The \emph{local} observer would be
wondering what causes this ``unnatural" explosion, which, at the very last moment, avoided the formation of a permanent
black hole; in our theory, however, the local observer would see no reason to use this function \(\w\), so she would not
notice any causality violation. Only the distant, outside observer would use this \(\w\), concluding that, indeed, the black
hole is not an eternal one since it evaporates.

The metric described in Figure~\ref{inoutmetric.fig} is related to the metrics described in Ref.~\cite{DraytH}. There, care
was taken that no conical singularity should arise when ingoing and outgoing shells meet. Here, we do allow this
singularity, interpreting it as a region of large \(R\) values, which we now demand not to be locally observable.

\newsecl{Conclusions}{conclusions}
Consider a large region of space-time, filled with light particles scattered here and there. Suppose we ask what this state
looks like after a very large Lorentz boost. This question is significant for instance if we consider the neighborhood of a
black hole horizon; a (large) time boost for an external observer corresponds to a (large) Lorentz boost for a local
observer. A problem with that is that light particles in one frame transform to extremely energetic ones in the boosted
frame. Their energies may become so large that the ensuing back reaction upon the metric may no longer be ignored. As we saw
in Section~\ref{scaleinv}, the effects on the space-time metric due to light particles moving nearly with the speed of light
can be efficiently represented by a conformal factor in the metric\fn{The configurations described in Section~\ref{scaleinv}
are actually limited to superpositions of flat shells of massless matter; this would leave a problem for massless particles
that are point-like. In a quantum theory this happens to be not such a bad restriction, if we assume that the particles form
plane waves, but we will not further expand on this issue here.}. This means that large Lorentz transformations may work as
usual only on spaces with a \(\hat g_{\mu\nu}\) metric, but generate delicate gravitational corrections in the scale
function \(\w(x)\). This may be seen as a different way to phrase our motivation for claiming that \(\w(x)\) should be
considered as locally unobservable, in particular when black hole horizons are considered. This way, the Lorentz group can
be kept as a perfect symmetry near the Planck scale, but only for the \(\hat g_{\mu\nu}\) part of the metric.

Our theory avoids the need to treat space-time as ``emergent"\cite{emergent}\cite{graphiti}. Rather, we concentrate upon the
demands of causality and locality. The light cones therefore come first, and the scale function \(\w\) is of secondary
importance.

Describing matter in a \(\hat g_{\mu\nu}\) metric will still be possible as long as we restrict ourselves to conformally
invariant field theories, which may perhaps be not such a bad constraint when describing physics at the Planck scale. Of
course that leaves us the question where Nature's mass terms come from, but an even more urgent problem is to find the
equations for the gravitational field itself, considering the fact that Newton's constant \(G_N\) is not scale-invariant at
all. The Einstein-Hilbert action is not scale-invariant. Here, we cannot use the Riemann curvature or its Ricci components,
but the Weyl component is independent of \(\w\), so that may somehow have to be used.

We suspect that, eventually, scales enter into our world in the following way. Information is now strictly limited to move
along the light cones, since only light like geodesics are well-defined, not the time like or space like ones. It is
generally believed that the amount of information moving around in Nature is limited to exactly one bit in each surface
element of size \(4\,\ln 2\) Planck lengths squared. Turning this observation around, one might assume that, whatever the
equations are, they define information to flow around. The density of this information flow may well define the Planck
length locally, and with that all scales in Nature. Obviously, this leaves us with the problem of defining what exactly
information is, and how it links with the equations of motion. The notion of information might not be observer-independent,
as the scale factor \(\w\) isn't. Quantum mechanics will probably require that all these bits of information form distinct
elements of a basis for Hilbert space, as described in Refs.~\cite{GtHdeterm}\cite{GtHentangled}.

\end{document}